\documentclass[aps,twocolumn,showpacs,preprintnumbers]{revtex4}
\usepackage{amssymb}
\usepackage{amsmath}
\usepackage{t1enc}
\usepackage{graphicx}

\begin{document}

\title{Mapping a quantum state of light onto a\\
long-lived atomic spin state: towards quantum memory}
\author{C. Schori}
\author{B. Julsgaard}
\author{J. L. Sørensen}
\author{E. S. Polzik}
\email{polzik@ifa.au.dk}
\affiliation{QUANTOP - Danish Quantum Optics Center\\
Institute of Physics and Astronomy, University of Aarhus, 8000 Aarhus,
Denmark}
\date{\today }

\begin{abstract}
We report an experiment on mapping a quantum state of light onto the
ground state spin of an ensemble of Cs atoms with the life time of 2
milliseconds.  Quantum memory for one of the two quadrature phase
operators of light is demonstrated with vacuum and squeezed states of
light. The sensitivity of the mapping procedure at the level of
approximately one photon/sec per Hz is shown. The results pave the
road towards complete (storing both quadrature phase observables)
quantum memory for Gaussian states of light. The experiment also sheds
new light on fundamental limits of sensitivity of the magneto-optical
resonance method.
\end{abstract}

\maketitle

\hyphenation{sta-te} 
\hyphenation{atoms} 
\hyphenation{two} 
\hyphenation{its}
\hyphenation{via} 
\hyphenation{mode} 
\hyphenation{con-firm}
\hyphenation{re-fers} 
\hyphenation{do-mi-nant}
\hyphenation{sta-ti-stics}
\hyphenation{e-qua-tions}
\hyphenation{Sto-kes}
\hyphenation{fluc-tua-tions}
\hyphenation{va-lue}
\hyphenation{Wals-worth}

\renewcommand{\S}[1]{\hat{S}_{#1}} 
\renewcommand{\a}[1]{{\hat{a}}_{#1}}
\renewcommand{\b}[1]{{\hat{b}}_{#1}}
\renewcommand{\vec}[1]{\mathbf{#1}}
\renewcommand{\H}[0]{\hat{H}}

Quantum state exchange between light and atoms is an important
ingredient for the future quantum information networks. It is also
crucial for sensitive atomic measurements in optics when quantum
limits of accuracy are approached. \ As shown recently both
theoretically \cite{kozhekin:00,kuzmich:97} and experimentally
\cite{hald:99,hald:00}, for ensembles of atoms such exchange is
feasible via free space interaction with light, as opposed to the case
of a single atom which requires cavity QED settings for that purpose
\cite{briegel:99}. In \cite{hald:99,hald:00} a short-lived squeezed
spin state of an atomic ensemble has been generated via complete
absorption of non-classical light. An alternative approach to mapping
a quantum state of light onto an atomic state via electromagnetically
induced transparency has been proposed \cite{fleischhauer:00} and
first experiments showing the feasibility of the method for classical
amplitude and phase of light have been carried out
\cite{phillips:01,liu:01}.

In this Letter we demonstrate for the first time a possibility of
mapping a quantum state of light onto a long-lived atomic spin
state. A novel approach utilizing a measurement induced back-action
along the lines of the proposal \cite{kuzmich:00b} is applied. We make
use of the strong non-dissipative off-resonant coupling between the
collective atomic spin and the polarization state of light. Such
coupling has been recently used for generation of a spin squeezed
atomic sample \cite{kuzmich:00a} and for entanglement of two separate
atomic objects \cite{julsgaard:01}. The coupling yields partial
information about the ground atomic spin state via a measurement
performed on the transmitted probe light. However, simultaneously via
the same interaction the quantum state of the optical probe is mapped
onto the conjugate atomic spin component which therefore serves as a
quantum memory.

We consider an ensemble of atoms with the collective spin
$\vec{J}=\sum \vec{j}^{(i)}$ where $\vec{j}^{(i)}$ is the total
angular momentum of the $i$'th atom in the ground state. Throughout
this paper the atoms are assumed to be spin polarized along the
$x$-direction, i.e.~$J_{x}$ is classical. The transverse spin
components satisfy $\left[ {\hat{J}}_{y},{\hat{J}}_{z}\right]
=iJ_{x}$. The experiment is carried out with cesium atoms in the $F=4$
hyperfine ground state.

\begin{figure}[t]
\includegraphics[width=\linewidth]{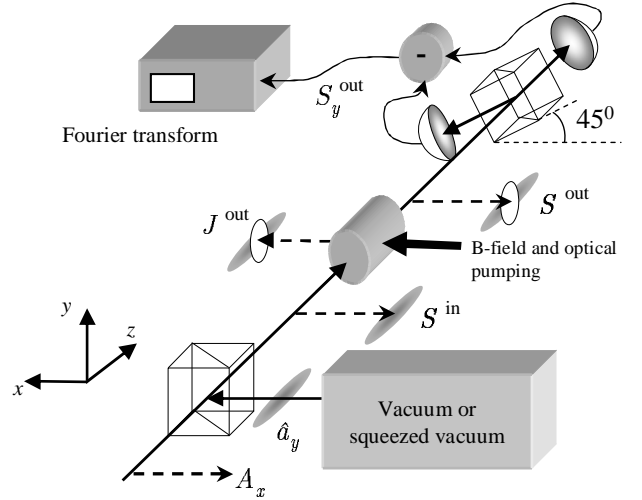}
\caption{A vacuum or squeezed vacuum field propagates through the
atomic sample, and leaves its trace on the sample. The transmitted
light which reads out this trace is analyzed at a polarization state
analyzer.}
\label{fig1}
\end{figure}

We study mapping and storage of Gaussian states of light, more
precisely, a vacuum or a squeezed vacuum state. This light described
by the continuum mode $\a{y}$ \cite{blow:90} is taken to be linearly
polarized along the $y$ -direction. In order to enable the free space
quantum state exchange with atoms, $\a{y}$ is mixed on a polarizing
beam splitter (Fig.~\ref{fig1}) with a strong classical field $A_{x}$
polarized along the $x$-direction. The polarization state of light
after the beam splitter is described by Stokes parameters
$\S{x}=\frac{1}{2}\left( A_{x}A_{x}-\a{y}^{\dagger }\a{y}\right)
=\frac{1}{2}A_{x}^{2}$ , $\S {y}=\frac{1}{2}\left( A_{x}\a{y}+\hat{a}
_{y}^{\dagger }A_{x}\right) =\frac{A_{x}}{2}\left(
\a{y}+\a{y}^{\dagger }\right) $ , $\S {z}=\frac{1}{2i}\left(
A_{x}\a{y}-\a{y}^{\dagger }A_{x}\right) =\frac{A_{x}}{2i}\left(
\a{y}-\hat{a}_{y}^{\dagger }\right)$
normalized to have the dimension sec$^{-1}$. In the experiment we
measure $2\S{y}$ which is the difference between the photon fluxes
polarized along $\pm 45^{\circ }$ directions.  Operators $\a{i}$ and
$\hat{a}_{j}^{\dagger }$ fulfill $\left[ \a{i}(\omega
),\hat{a}_{j}^{\dagger }(-\omega ^{\prime })\right] =\delta (\omega
-\omega ^{\prime })\delta _{ij}$. With $S_{x}$ regarded as classical,
correlation functions for the Stokes operators are $\left\langle \S
{i}(\omega )\S {j}(-\omega ^{\prime })\right\rangle =\frac{
S_{x}}{2}\epsilon _{i}\delta (\omega -\omega ^{\prime })\delta _{ij}$
with $i,j=y,z$. The Stokes operators $\S {y}$ or $\S{z}$ are squeezed
if the $\a{y}$ mode is in a squeezed state, and $\epsilon_i \lessgtr 1$
refers to squeezing/anti-squeezing. With a coherent probe $\epsilon
_{y}=\epsilon _{z}=1$.

The interaction between the light and the atoms is modeled following
\cite{kuzmich:00b}, \cite{duan:00a} to obtain
\begin{equation}
\label{eq:Sout}
\S {z}^{\text{out}}(t) =\S {z}^{\text{in}}(t),\quad \S {y}^{\text{out}}(t) =
\S {y}^{\text{in}}(t)+aS_{x}{\hat{J}}_{z}(t)
\end{equation}
where \textquotedblleft in\textquotedblright~and \textquotedblleft
out\textquotedblright~refer to the light before and after interaction
with the atoms. These equations are valid if the light is sufficiently
far detuned from the atomic resonance, so that polarization
rotation from circular birefringence is the dominant contribution
to the interaction. Note that this rotation does not affect
$\S{z}$ but can be read out in $\S{y}$. The strength of the
interaction is described by the parameter $a$ which depends on the
beam geometry, detuning and the interaction cross section
\cite{kuzmich:00b}.

The evolution of the atomic spins is described by the equations 
\begin{align}
\dot{\hat{J}}_{z}(t)& =\Omega {\hat{J}}_{y}-\Gamma {\hat{J}}_{z}(t)+{\hat{
\mathcal{F}}}_{z}(t)  \label{JzDot} \\
\dot{\hat{J}}_{y}(t)& =-\Omega {\hat{J}}_{z}-\Gamma {\hat{J}}_{y}(t)+{\hat{
\mathcal{F}}}_{y}(t)+aJ_{x}\S {z}(t)
\label{JyDot}
\end{align}
where, as in \cite{julsgaard:01}, we introduce a constant magnetic
field along the $x$-direction giving rise to Larmor precession with
the frequency $\Omega$. In what follows, the value of $\Omega$
determines the frequency component of light which can be stored in the
atomic medium. The decay of the transverse ground state spin
components is described by the rate $\Gamma$ and their quantum
statistics are described by the Langevin forces
${\hat{\mathcal{F}}}_{y}$, ${\hat{\mathcal{F}}}_{z}$. The term
$aS_{x}{\hat{J}}_{z}(t)$ in the equation~(\ref{eq:Sout}) is
responsible for the spin state read out. The term $aJ_{x}\S {z}(t)$ in
equation~(\ref{JyDot}) is the so-called back action of light onto
atoms. This latter term feeds the quantum fluctuations of light into
atoms. Taking the Fourier transforms of equations~(\ref{JzDot})
and~(\ref{JyDot}) we obtain
\begin{align}
{\hat{J}}_{y}(\omega )& =-\frac{1}{2}\cdot \frac{iaJ_{x}\S {z}(\omega )+i{
\hat{\mathcal{F}}}_{y}(\omega )+{\hat{\mathcal{F}}}_{z}(\omega )}{(\Omega
-\omega )-i\Gamma }  \label{eq:JzOmega} \\
{\hat{J}}_{z}(\omega )& =\frac{1}{2}\cdot \frac{aJ_{x}\S {z}(\omega )+{\hat{
\mathcal{F}}}_{y}(\omega )-i{\hat{\mathcal{F}}}_{z}(\omega )}{(\Omega
-\omega )-i\Gamma }
\end{align}
where we have made the narrow-band approximation $|\omega -\Omega |\ll
\Omega $ and $\Gamma \ll \Omega $. Combining~(\ref{eq:JzOmega}) with the
Fourier transform of~(\ref{eq:Sout}) we get $\S {y}^{\text{out}}(\omega )=\S 
{y}^{\text{in}}(\omega )+\frac{\frac{1}{2}aS_{x}}{(\Omega -\omega )-i\Gamma }
\left\{ aJ_{x}\S {z}(\omega )+{\hat{\mathcal{F}}}_{y}(\omega )-i{\hat{
\mathcal{F}}}_{z}(\omega )\right\} $. To calculate the power spectrum of $\S 
{y}^{\text{out}}$, we must know the correlation function for the Langevin
noise terms. The role of these terms is to preserve the correct commutation
relations for ${\hat{J}}_{y}$ and ${\hat{J}}_{z}$ in the presence of the
decoherence described by $\Gamma $. In the experiment the decoherence is
primarily caused by a resonant optical pump laser driving the atoms into the
coherent spin state $F=4$, $m_{F}=4$ \cite{kitagawa:93}, and, to a much 
smaller extent, by collisions and quadratic Zeeman effect.

In order to compute the noise correlation functions we integrate
equation~(\ref{JzDot}) omitting the magnetic field and optical probe
terms which do not contribute to the decoherence, i.e.~with $\Omega
=0$ and $a=0.$ The result is
\begin{equation}
\left\langle {\hat{J}}_{z}(t)^{2}\right\rangle =\left\langle {\hat{J}}
_{z}(0)^{2}\right\rangle e^{-2\Gamma t}+\frac{k_{zz}}{2\Gamma }\left(
1-e^{-2\Gamma t}\right) 
\end{equation}
where we have assumed a memory-less reservoir \cite{gardiner:00} of
the form
$\left\langle{\hat{\mathcal{F}}}_{z}(t){\hat{\mathcal{F}}}_{z}(t^{\prime
})\right\rangle =k_{zz}\delta (t-t^{\prime })$. Using the coherent
spin state variance, $\left\langle {\hat{J}}_{z}^{2}\right\rangle
=J_{x}/2,$ we obtain $k_{zz}=\Gamma J_{x}$. The Fourier-transformed
correlation function can then easily be shown to fulfill $\left\langle
{\hat{\mathcal{F}}} _{z}(\omega ){\hat{\mathcal{F}}}_{z}^{\ast
}(\omega ^{\prime })\right\rangle =J_{x}\Gamma \delta (\omega -\omega
^{\prime })$. With similar equations for the $y$-components we end up
with the expression for the power spectrum $\Phi(\omega)$ of
$\S{y}^{\text{out}}$ observed in the experiment
\begin{equation}
\Phi(\omega ) =\frac{S_{x}}{2}
\epsilon _{y}+\frac{\frac{1}{4}a^{2}S_{x}^{2}}{(\Omega -\omega )^{2}+\Gamma
^{2}}\left\{ \frac{a^{2}J_{x}^{2}S_{x}\epsilon _{z}}{2}+2\Gamma
J_{x}\right\}   \label{eq:spectrum_out}
\end{equation}
The first term is determined by the $\S {y}$ component of the input
light, i.e.~by the quadrature phase operator $\frac{1}{2}\left(
\a{y}+\hat{a}_{y}^{\dagger }\right) $. The rest is due to the narrow
band atomic state fluctuations with the life time $\Gamma ^{-1}$. The
first term in the curly brackets is the back action noise due to the
quantum state of the input light, more precisely due to the operator
$\frac{1}{2i}\left( \a{y}-\hat{a}_{y}^{\dagger }\right) $. This is the
"quantum memory term". The second term in the brackets is the
projection noise of the initial atomic spin state.

\bigskip Turning now to the experiment, we achieve nontrivial
$\epsilon_{y,z}$ values by using squeezed vacuum light generated by
the frequency tunable degenerate optical parametric amplifier below
threshold \cite{polzik:98}. This light is overlapped with the
orthogonally polarized strong (up to 5mW) beam on a polarizing beam
splitter (see Fig.~\ref{fig1}). As shown in \cite{hald:00} we can, by
using this technique, vary the quantum fluctuations in $\S {y}$ to be
below or above the coherent state limit depending on the relative
phase between squeezed vacuum and the classical field. Typically the
degree of squeezing emerging from our source is about -5dB, but due to
various imperfections and losses only -3dB of Stokes parameter
squeezing is detected, corresponding to $\epsilon _{y}=0.5$. The
degree of anti-squeezing is typically 8-9dB relative to the coherent
state probe, corresponding to $\epsilon _{z}=6-8$.

Our atomic sample is Cs vapor in a paraffin coated glass cell. The
atomic angular momentum is created by optical pumping along the
$x$-axis with two $\sigma _{+}$ polarized diode lasers resonant with
the $6S_{1/2}(F=4)\rightarrow 6P_{1/2}(F=4)$ transition and the
$6S_{1/2}(F=3)\rightarrow 6P_{3/2}(F=4)$ transition. By adjusting the
relative power of the lasers we are able to control the number of
atoms in the $F=4$ ground state. The degree of spin polarization
($\approx 95$\%) and the number of atoms is measured by observing the
magneto-optical resonance of the ground spin state.

The output Stokes parameter $\S {y}$ is measured by a polarizing beam
splitter oriented at a 45 degree angle with respect to the mean input
optical polarization. The probe is blue detuned by 875MHz from the
$6S_{1/2}(F=4)\rightarrow 6P_{3/2}(F=5)$ transition of Cs atoms at
rest. The power spectrum of $\S {y}$ is recorded in a frequency window
varying from 1.6kHz to 3.2kHz around $\Omega$. The resulting spectrum
is a narrow Lorentzian centered at $\Omega$ with a width
$\Gamma$. This width can be varied by adjusting the power of the
optical pumping lasers with line widths in the range from 100Hz to
1kHz FWHM.

\begin{figure}[t]
\includegraphics[width=\linewidth]{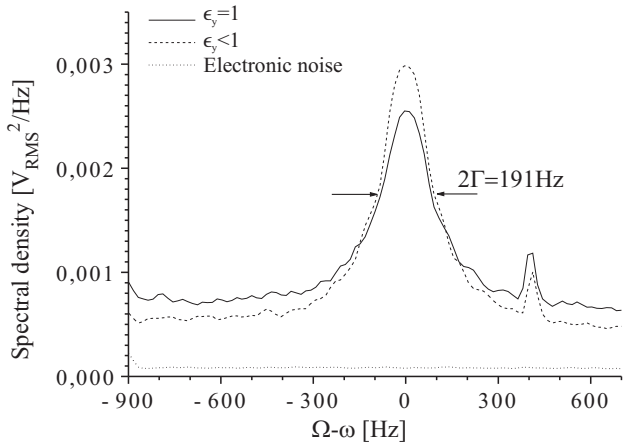}
\caption{The measured spectrum of the transmitted probe. The solid
line is obtained with the input light in a vacuum state
($\protect\epsilon_y=\protect\epsilon_z=1)$. When the input mode is
in a squeezed state (dashed line) the Lorentzian part from atoms
increases while the wings decrease. The peak on the right is technical
noise.}
\label{fig2}
\end{figure}

\begin{figure}[t]
\includegraphics[width=\linewidth]{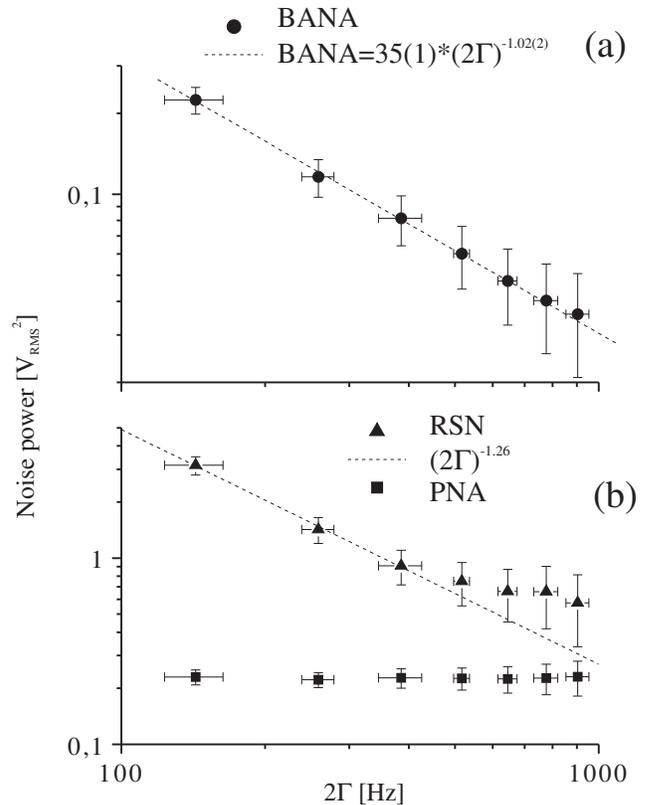}
\caption{\textbf{(a)} The measured back action noise area (BANA) for
the vacuum light input as a function of decay rate $\Gamma$ on a
log-log scale.  \textbf{(b)} The measured residual spin noise (RSN)
and the inferred projection noise area (PNA) calculated
from~(\protect\ref{eq:PNA}). See text for details.}
\label{fig3}
\end{figure}

A typical spectrum is shown in Fig.~\ref{fig2}. This spin noise
resonance contains contributions from the spin projection noise, the
back action term and the technical noise of the spin. The latter is
not known, however, as shown below, it can be evaluated, and the
consistency of the model can be verified by varying the quantum state
of the probe. The two upper traces in the figure correspond to the
vacuum state of the mode $\a{y}$ (the probe in a coherent polarization
state) and the squeezed vacuum state of $\a{y}$ (the probe in a
squeezed polarization state), respectively. The reduction of the
optical quantum noise of $\S {y}$ is clearly seen in the wings of the
atomic resonance when the polarization squeezed probe is
applied. However, more interesting is that the atomic spin noise
\emph{grows} when the probe is squeezed in $\S{y}$. This is due to the
light-induced back action noise of the atoms. Note that this back
action may completely wash out the advantages of using squeezed light
in sensitive magnetometry \cite{kupriyanov:92}.

Figure~\ref{fig2} represents the evidence of anti-squeezed Stokes
parameter $\S{z}$ being stored in the atoms. This storage takes place
over a time scale on the order of ($2\pi \Gamma )^{-1}$ which here
corresponds to about 2 milliseconds. Note that squeezed vacuum output
of the optical parametric amplifier \cite{polzik:98} at our values of
the gain contains about one photon per second per Hz of the
bandwidth. Hence Fig.~\ref{fig2} shows the effect of approximately 200
photons/sec stored within the atomic bandwidth. We may expect
therefore the sensitivity of the atomic memory at the level of a
single photon for pulses with $(2\pi\Gamma)^{-1}\approx 2$ms duration.

From eq.~(\ref{eq:spectrum_out}) we know that the \emph{back action
noise area} (BANA) is proportional to $\epsilon _{z}$, so that we can
extract this contribution from the overall atomic spin noise by
$\mbox{BANA}=(A_{SQ}-A_{COH})/(\epsilon _{z}-1)$, where $A_{COH}$ and
$A_{SQ\;}$ are the measured spin noise areas with a coherent and
squeezed probe, respectively.  We can also define the \emph{residual
spin noise}
$\mbox{RSN}=(\epsilon_{z}A_{COH}-A_{SQ})/(\epsilon_{z}-1)$.

To compare the experimental results with theoretical predictions we
integrate (\ref{eq:spectrum_out}) over frequencies and calculate the back
action noise area 
\begin{equation}  
\label{eq:BANA}
\text{BANA}=\frac{\pi a^{4}J_{x}^{2}\epsilon _{z}}{\Gamma }
\left( \frac{S_{x}}{2}\right) ^{3}
\end{equation}
and the \emph{projection noise area} 
\begin{equation}  \label{eq:PNA}
\text{PNA}=2\pi a^{2}J_{x}\left( \frac{S_{x}}{2}\right) ^{2}
\end{equation}
Knowing the \emph{shot noise level} $\text{SNL}=S_{x}/2$ we obtain for a
coherent probe ($\epsilon _{y}=\epsilon _{z}=1$), 
\begin{equation}
\text{PNA}=2\sqrt{\pi \Gamma (\text{BANA})\times (\text{SNL})}
\label{eq:noise_depend}
\end{equation}
Varying $\epsilon _{z}$ we can determine the back action term~(\ref{eq:BANA}
), which is in turn determined by the quantum state of the probe light. With
the knowledge of the BANA and the SNL, we are able to deduce the size of the
PNA from equation~(\ref{eq:noise_depend}).

\begin{figure}[t]
\includegraphics[width=\linewidth]{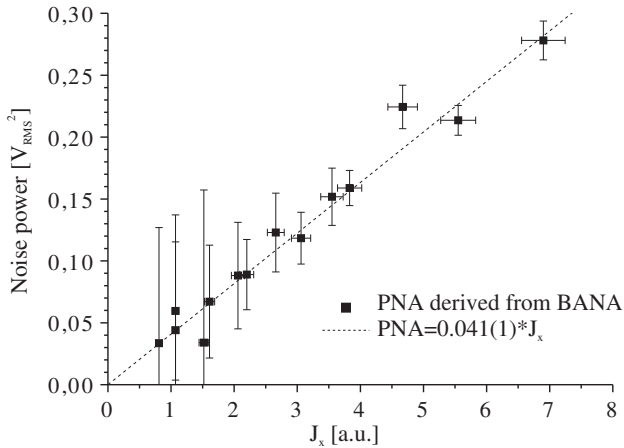}
\caption{The projection noise area derived
from~(\protect\ref{eq:noise_depend}). The data is, in fact, compiled
from two series with different $\Gamma$'s, giving strong quantitative
support of the model.}
\label{fig4}
\end{figure}

We now vary the probe power to confirm the $S_{x}^{3}$ dependence of
the back action noise and the number of atoms to confirm the
$J_{x}^{2}$ dependence. We find the BANA to scale with $S_{x}^{p}$,
where $p=2.8\pm 0.2 $, and $J_{x}^{q}$ with $q=2.0\pm 0.2$ in good
agreement with the theory.  During the sequence of measurements
$\epsilon _{y}$ and $\epsilon _{z}$ are monitored several times and
variations are found to be less than 10\%. Even more exciting is to
observe the variation of BANA and RSN with $\Gamma$. From
Fig.~\ref{fig3}(a) we see that the BANA decreases as expected with
$\Gamma ^{-1}$ to within a few percent. The RSN dependence on $\Gamma
$, shown in Fig.~\ref{fig3}(b), is found to fall off roughly with
$\Gamma ^{-1}$ until a value of approximately $2\Gamma =500$Hz is
reached.  Here the RSN starts leveling out to approach a constant
value. This is due to the higher degree of optical pumping, and hence
higher $\Gamma $, which pushes the atomic spin towards the coherent
spin state (CSS) and thereby washes out any additional (technical)
noise in the spin state. We therefore expect the RSN to converge to a
level set by the CSS. In our notation this level is just the PNA, and
using eq.~(\ref{eq:noise_depend}) we can estimate this level from the
BANA. This estimate is also shown in Fig.~\ref{fig3}(b) where we see
that the RSN indeed is approaching the PNA at higher values of
$\Gamma$.

Finally, we fix $\Gamma$ and vary the number of atoms ($J_{x}$), see
Fig.~\ref{fig4}. Again from the measured BANA we can infer the PNA and
observe that it grows proportionally with $J_{x}$ to test of the
consistency of our model. The plotted results are obtained at two very
different decay rates $2\Gamma =264\pm 20$ Hz and $2\Gamma =485\pm
20$Hz, but nonetheless fit nicely to the same line confirming the
theory.

In summary we have shown how partial information about an unknown
Gaussian quantum state of light, more precisely the value of the
operator $\frac{1}{2i}\left( \a{y}-\hat{a}_{y}^{\dagger }\right)$, can
be mapped onto the atomic ground state spin, where it is stored for
approximately 2ms. We have also demonstrated how this information is
read out again by the probe via the narrow band atomic spin noise
around the Larmor frequency. Most importantly we have demonstrated
that long-lived atomic spin ensembles can serve as quantum memory for
light sensitive enough to store quantum states of optical fields
containing just a few photons.

In our continuous wave experiment the probe effectively reads out its
own quantum state at an earlier time which has been stored in
atoms. Future experiments using two pulses, one to be stored and
another to read out the stored information, are clearly
feasible. Another important step towards full scale quantum memory for
light will be to achieve storage and retrieval of the full quantum
state of light, which for a Gaussian state corresponds to both
$\frac{1}{2i}\left( \a{y}-\hat{a}_{y}^{\dagger }\right) $ and
$\frac{1}{2}\left( \a{y}+\hat{a}_{y}^{\dagger }\right)$ operators. In
order to store the full quantum state of light one has to use two
entangled atomic ensembles \cite{julsgaard:01}, rather than just one
ensemble, as in the present paper. For the atomic state read-out
\cite{kuzmich:00b} two entangled beams of light, rather than a single
beam, as in the present work, should be used.

The narrow band frequency response of the atomic ground state spin
allows only for storage of a few hundred Hz wide frequency band of the
optical state around the carrier frequency $\Omega $. One way to
overcome this limit would be to apply a linear magnetic field gradient
in the $z$-direction, or to use an inhomogeneously broadened solid
state medium \cite{turukhin:02}. This would make atoms at different
positions have different Larmor frequencies, hence they would store
different frequency components of the optical state and the entire
optical spectrum of interest could then be mapped onto atoms.

\begin{acknowledgments}
This work was supported by the Danish National Research Foundation and by
the EU QIPC network via the QUICOV project.
\end{acknowledgments}

\bibliographystyle{apsrev}
\bibliography{bibfile}

\end{document}